\documentclass[11pt,a4paper,showpacs,showkeys,superscriptaddress]{article}

\usepackage{epsfig}
\usepackage{amsmath,amssymb}
\usepackage{graphicx}
\usepackage{amstext}
\usepackage{mathrsfs}

\usepackage{latexsym}
\usepackage{hyperref}
\hypersetup{colorlinks,%
  citecolor=blue,%
  linkcolor=red,%
  pdftex}

\begin{document}
\baselineskip 6mm

\newcommand{\nc}{\newcommand}
\newcommand{\rnc}{\renewcommand}



\newcommand{\tcb}{\textcolor{blue}}
\newcommand{\tcr}{\textcolor{red}}
\newcommand{\tcg}{\textcolor{green}}


\def\beq{\begin{equation}}
\def\eeq{\end{equation}}
\def\ba{\begin{array}}
\def\ea{\end{array}}
\def\bea{\begin{eqnarray}}
\def\eea{\end{eqnarray}}
\def\nn{\nonumber}


\def\CMP{Commun. Math. Phys.~}
\def\JHEP{JHEP~}
\def\Pre{Preprint}
\def\PRL{Phys. Rev. Lett.~}
\def\PR {Phys. Rev.~}
\def\CQG {Class. Quant. Grav.~}
\def\PL {Phys. Lett.~}
\def\NP {Nucl. Phys.~}

\def\G{\Gamma}

\def\S{{\bf S}}
\def\C{{\bf C}}
\def\Z{{\bf Z}}
\def\R{{\bf R}}
\def\N{{\bf N}}
\def\M{{\bf M}}
\def\P{{\bf P}}
\def\bm{{\bf m}}
\def\bn{{\bf n}}

\def\CA{{\cal A}}
\def\CB{{\cal B}}
\def\CC{{\cal C}}
\def\CD{{\cal D}}
\def\CE{{\cal E}}
\def\CF{{\cal F}}
\def\CH{{\cal H}}
\def\CM{{\cal M}}
\def\CG{{\cal G}}
\def\CI{{\cal I}}
\def\CJ{{\cal J}}
\def\CL{{\cal L}}
\def\CK{{\cal K}}
\def\CN{{\cal N}}
\def\CO{{\cal O}}
\def\CP{{\cal P}}
\def\CQ{{\cal Q}}
\def\CR{{\cal R}}
\def\CS{{\cal S}}
\def\CT{{\cal T}}
\def\CU{{\cal U}}
\def\CV{{\cal V}}
\def\CW{{\cal W}}
\def\CX{{\cal X}}
\def\CY{{\cal Y}}
\def\CZ{{\cal Z}}

\def\We{{W_{\mbox{eff}}}}


\newcommand{\Lie}{\pounds}

\newcommand{\p}{\partial}
\newcommand{\bp}{\bar{\partial}}

\newcommand{\half}{\frac{1}{2}}

\newcommand{\bfalpha}{{\mbox{\boldmath $\alpha$}}}
\newcommand{\bfbeta}{{\mbox{\boldmath $\beta$}}}
\newcommand{\bfgamma}{{\mbox{\boldmath $\gamma$}}}
\newcommand{\bfmu}{{\mbox{\boldmath $\mu$}}}
\newcommand{\bfpi}{{\mbox{\boldmath $\pi$}}}
\newcommand{\bfvarpi}{{\mbox{\boldmath $\varpi$}}}
\newcommand{\bftau}{{\mbox{\boldmath $\tau$}}}
\newcommand{\bfeta}{{\mbox{\boldmath $\eta$}}}
\newcommand{\bfxi}{{\mbox{\boldmath $\xi$}}}
\newcommand{\bfkappa}{{\mbox{\boldmath $\kappa$}}}
\newcommand{\bfepsilon}{{\mbox{\boldmath $\epsilon$}}}
\newcommand{\bfTheta}{{\mbox{\boldmath $\Theta$}}}

\newcommand{\bz}{{\bar{z}}}

\newcommand{\dalpha}{\dot{\alpha}}
\newcommand{\dbeta}{\dot{\beta}}
\newcommand{\blambda}{\bar{\lambda}}
\newcommand{\btheta}{{\bar{\theta}}}
\newcommand{\bsigma}{{{\bar{\sigma}}}}
\newcommand{\bepsilon}{{\bar{\epsilon}}}
\newcommand{\bpsi}{{\bar{\psi}}}


\def\ct{\cite}
\def\la{\label}
\def\eq#1{(\ref{#1})}


\def\a{\alpha}
\def\b{\beta}
\def\g{\gamma}
\def\G{\Gamma}
\def\d{\delta}
\def\D{\Delta}
\def\ep{\epsilon}
\def\e{\eta}
\def\ph{\phi}
\def\Ph{\Phi}
\def\ps{\psi}
\def\Ps{\Psi}
\def\k{\kappa}
\def\l{\lambda}
\def\L{\Lambda}
\def\m{\mu}
\def\n{\nu}
\def\th{\theta}
\def\Th{\Theta}
\def\r{\rho}
\def\s{\sigma}
\def\S{\Sigma}
\def\ta{\tau}
\def\o{\omega}
\def\O{\Omega}
\def\pr{\prime}
\def\f{\varphi}


\def\half{\frac{1}{2}}

\def\goto{\rightarrow}

\def\na{\nabla}
\def\grad{\nabla}
\def\curl{\nabla\times}
\def\div{\nabla\cdot}
\def\pa{\partial}

\def\bra{\left\langle}
\def\ket{\right\rangle}
\def\lb{\left[}
\def\lc{\left\{}
\def\ls{\left(}
\def\lp{\left.}
\def\rp{\right.}
\def\rb{\right]}
\def\rc{\right\}}
\def\rs{\right)}
\def\cl{\mathcal{l}}

\def\vac#1{\mid #1 \rangle}

\def\td#1{\tilde{#1}}
\def\check{ \maltese {\bf Check!}}


\def\Tr{{\rm Tr}\,}
\def\det{{\rm det}\,}


\def\bc#1{\nnindent {\bf $\bullet$ #1} \\ }
\def\ch {$<Check!>$ }
\def\ss {\vspace{1.5cm}}

\begin{titlepage}
%
%
%
%
%
%
%
%
\begin{center}
{\Large \bf Universality of the Unruh effect}
%
\vskip 1. cm
  { ${}^{a}$Leonardo Modesto\footnote{e-mail : lmodesto@sustc.edu.cn},
   ${}^{b}$Yun Soo Myung\footnote{e-mail : ysmyung@inje.ac.kr},
 ${}^{c}$Sang-Heon Yi\footnote{e-mail : shyi704@uos.ac.kr}
 }
\vskip 0.5cm
{\it ${}^{a}$Department of Physics, Southern University of Science and Technology, Shenzhen 518055, China\\
${}^{b}$Institute of Basic Science and Department of Computer Simulation,
Inje University, Gimhae 50834, Korea\\
 ${}^{c}$Physics Department, University of Seoul, Seoul 02504, Korea }\\
\end{center}
\thispagestyle{empty}
\vskip1.5cm
%
%
\centerline{\bf ABSTRACT} \vskip 4mm
 \vspace{1cm}
\noindent
In this paper we prove the universal nature of the Unruh effect in a general class of weakly non-local field theories. At the same time we solve the tension between two conflicting claims published in literature. Our universality statement is based on two independent computations based on the canonical formulation as well as path integral formulation of the quantum theory.

\vspace{2cm}
%
%
\end{titlepage}
\renewcommand{\thefootnote}{\arabic{footnote}}
\setcounter{footnote}{0}
%
%
%
%

\section{Introduction}
  As  mentioned in the literature~\cite{Woodard:2006nt}, the non-locality  has its importance at various levels.  On the one hand, the non-locality may originate from a fundamental theory like string theory \cite{CaMo2} or from the quantum spacetime~\cite{Bombelli:1987aa}.  On the other hand, the non-locality may come out as the low-energy phenomena after integrating out the higher momentum degrees of freedom, just as realized in the quantum effective action. Finally, the nature could be intrinsically non-local and the non-local field theory will turn out to be the most appropriate framework to describe it \cite{krasnikov,kuzmin,tom,Modesto:2011kw}.
In all these cases, non-local theories urge us to revisit the well-established results found in local field theory. In fact, many papers have followed  this route in non-commutative field theories (see~\cite{Douglas:2001ba} for a review), which could be regarded as a kind of non-local theories.

One of the recent renewed interests in non-local theories is due to a class of  super-renormalizable or finite quantum gravity models (see for examples,~\cite{Modesto:2011kw,Modesto:2014lga}). These non-local theories have improved behaviour in the ultraviolet  regime and it can be proved on the base of rigorous power counting arguments that there is class of non-local gravitational theories super-renormalizable at quantum level.
The good behaviour at quantum level and the perturbative unitarity
provide us a good motivation to investigate non-local theories.
Contrary to the non-commutative field theory, these theories preserve the Lorentz symmetry
and do not violate the macroscopic causality.  Here, the non-locality may be represented  by a length scale $\ell$ and one may ask  what is the effect of the non-local scale $\ell$ in various settings.

To see whether there are non-local effects in the quantum field theory setup, we here focus on a toy model presenting non-local modifications only on the field theory propagator, while the interactions are the usual local ones.
Therefore, one can ask whether the non-locality and the related scale $\ell$ do affect the results of the free field theory quantization. One of those effects is the Unruh one~\cite{Unruh:1976db} seen by a uniformly accelerating observer. Recently, in the non-local theory, whose non-locality is given by a particular entire function, the Unruh effect appeared with opposite outcomes: a significant modification in~\cite{Nicolini:2009dr} and no modification~\cite{Kajuri:2017jmy}. In order to resolve the tension between the two results, we here revisit the Unruh effect in a non-local theory of ``Gaussian exponential type". However, the result applies to any weakly non-local field theory.

This paper is organized as follows. In section two, we briefly review the Unruh effect by employing the canonical quantization for a specific non-local model. In section three, we revisit the Unruh-DeWitt detector method for the Unruh effect and we show  the irrelevance of the non-locality scale $\ell$ to the Unruh effect. In section four, we confirm our results by implementing a field redefinition in the Lagrangian of the theory.  We summarize our results in the section five.

\section{Review:  The canonical method}
The Unruh effect is well-established as the quantum phenomenon seen by a  uniformly accelerating (Rindler) observer. The Rindler observer confined in the Rindler wedge of the whole Minkowski spacetime  cannot access the other parts of Minkowski spacetime and so, he/she has to consider the Minkowski vacuum as  an entangled mixed state. After computation, it turns out that the Minkowski vacuum could be regarded as a thermal state with its temperature $T_{\text{DU}}$ known as the Davies-Unruh (DU) temperature. The DU temperature is given  by   the proper acceleration $a$  of the observer or the detector as
\[
T_{\text{DU}} = \frac{a\hbar}{2\pi k_{B}}\,.
\]
Here, we will set $c=\hbar=k_{B}=1$  as usual and take the signature as $\eta_{\mu\nu}= (-,+,+,+)$.  One way to derive  $T_{\text{DU}}$ is to use the canonical quantization and adopt the  Boguliubov transformation between the Minkowski Fock space and the Rindler Fock space~\cite{Fulling:1972md}. In the following, we adopt the argument given in~\cite{Kajuri:2017jmy}  in order to fix our conventions and viewpoints. We would like to  mention some potential loopholes appeared in this approach.

Let us consider the simplest non-local field theory given by the following kinetic Lagrangian operator with gaussian form factor,
\begin{equation} \label{Our}
{\cal L} = -\frac{1}{2}\phi(x)\,  e^{-\frac{\ell^{2}}{2}\Box_{x}}(-\Box_{x})\,  \phi (x)\,.
\end{equation}
Then, the field equation could be written as
\begin{equation} \label{EOM}
e^{-\frac{\ell^{2}}{2}\Box_{x}}(-\Box_{x})\,  \phi (x) =0\,.
\end{equation}
Note that the conventional local case can be recovered  by taking the non-local scale $\ell$ to be zero.
Hereafter, we denote  the local scalar theory by introducing  the super/sub-script $(0)$ as
\begin{equation} \label{LST}
{\cal L}_{(0)} = -\frac{1}{2}\phi(x)\,  (-\Box_{x})\,  \phi (x)\,.
\end{equation}

In order to investigate the Unruh effect by the Bogoliubov transformation, it is necessary  to adopt the canonical quantization in the Minkowski and the Rindler spaces, respectively. To proceed in this way, we quantize  the the Lagrangian (\ref{Our}) canonically.  As was emphasized in~\cite{Kajuri:2017jmy},   the  non-local entire function  does not change the pole structure of the propagator at perturbative level~\cite{Barnaby:2007ve, Calcagni} and thus, the homogenous solution  $\phi(x)$ to the equation of motion (\ref{EOM}) is unique and given by
\begin{equation} \label{}
\phi(x) = \int d^{3}\pmb{p}\Big[ a_{\pmb{p}}u_{\pmb{p}} (x) + a^{\dagger}_{\pmb{p}}u^{*}_{\pmb{p}} (x)\Big]\,.
\end{equation}
At the classical level, the coefficients of $a_{\pmb{p}}$ and $a^{\dagger}_{\pmb{p}}$ correspond to the initial data of  $\phi$ to determine the evolution of the system. 

Now, one may  adopt the canonical quantization approach taken in~\cite{Barnaby:2007ve,Barci:1995ad,Aslanbeigi:2014zva}. This approach provides the  simple quantization rule
\begin{equation} \label{COMM}
[a_{\pmb{p}}, a^{\dagger}_{\pmb{p}'}] = \delta^{(3)}(\pmb{p}-\pmb{p}')\,, \qquad [a_{\pmb{p}}, a_{\pmb{p}'}] =[a^{\dagger}_{\pmb{p}}, a^{\dagger}_{\pmb{p}'}] =0\,.
\end{equation}
We admit that this quantization might not have the clear meaning as in the local theory, since  the equal-time commutator of the field and its momentum, which was used to derive Eq.~(\ref{COMM}),  is not well-defined in the non-local theory. Moreover, the non-locality does not allow us to convert easily the relation between momentum and velocity, which  gives rise to difficulties when  the Lagrangian is Legendre-transformed  to the canonical Hamiltonian. However, as was done in~\cite{Barci:1995ad}, one may take the canonical quantization without introducing the equal-time commutators and the canonical Hamiltonian\footnote{In Ref.~\cite{Barci:1995ad}, the Hamiltonian was read off from the time component of energy-momentum tensor, not through the Legendre transformation.}. Moreover, one may consider  (\ref{COMM})  as the fundamental quantization rules by analog with those  in the local theory~\cite{Weinberg:1995mt,Wald:1995yp}.  These quantization rules are also acceptable from the viewpoint that  the position-space non-locality in~(\ref{Our})  is ``local" in the momentum space. This means that  the non-locality in the position space  is just the UV modification of the propagator in the momentum space.

As explained in~\cite{Takagi:1986kn,Crispino:2007eb}, the Unruh effect may be understood through the Bogoliubov transformation of  annihilation/creation operators in  between  Minkowski and Rindler spaces.  Considering   arguments stated in~\cite{Kajuri:2017jmy},  the unchange  of  the Unruh effect or the DU temperature originates  from  the  commutation relations in~(\ref{COMM}).  For instance, the Klein-Gordon inner product for  $\phi(x)$'s can be defined as usual without any modifications and so all steps for the Bogoliubov transformation are  identical with those in the local theory.

\section{The Unruh-DeWitt detector method}
In this section, we revisit the Unruh-DeWitt detector approach for the model of (\ref{Our}). The Unruh-DeWitt detector~\cite{Unruh:1983ms}  is a uniformly accelerating hypothetical detector  with two  energy levels and its interaction with the surrounding field is given by
\begin{equation} \label{Lint}
\mathcal{L}_{\rm int} = g M(x(\tau)) \phi(x(\tau))\,,
\end{equation}
where $\tau$ denotes the proper time of the uniformly accelerating detector.

In the  detector approach to the Unruh effect, a physical quantity to compute is the  response function  $F(\omega)$ defined  by
\begin{equation} \label{Res}
F(\omega) = \int^{\infty}_{-\infty}d\tau \int^{\infty}_{-\infty}d \tau'~ e^{-i(\tau-\tau')} \langle 0_{M}| \phi(x(\tau))\phi(x(\tau'))|0_{M} \rangle\,,
\end{equation}
where $|0_{M} \rangle $ denotes the Minkowski vacuum. Here, we  assumed that the Minkowski vacuum can be defined through the canonical quantization.

At this stage, we  emphasize that the two-point function  in~(\ref{Res}) is not the Feynman propagator of  $\Delta_{F}(x,x')$,  but the positive-frequency  Wightman function $G^{+}(x,x')$. Note that  these two functions satisfy the inhomogeneous and homogeneous  equations, respectively, as
\begin{equation} \label{requirement1}
e^{-\frac{\ell^{2}}{2}\Box_{x}}(-\Box_{x})\Delta_{F}(x,x') = \delta^{4}(x-x')\,, \qquad e^{-\frac{\ell^{2}}{2}\Box_{x}}(-\Box_{x})G^{+}(x,x') =0\,,
\end{equation}
while both preserve translational symmetry as
\[
\Delta_{F}(x,x') = \Delta_{F}(x-x') \quad {\rm and} \quad G^{+}(x,x') = G^{+}(x-x').
\]
A big  difference between two equations in (\ref{requirement1}) is  the presence or not of the
source $ \delta^{4}(x-x')$.
Another distinguishing property between $\Delta_{F}$ and $G^+$ is
\begin{equation} \label{requirement2}
\Delta_{F}(-x) = \Delta_{F}(x)\,, \qquad G^{+}(-x) = G^{-}(x)\neq G^{+}(x)\,,
\end{equation}
where $G^{-}(x)$ is the negative-frequency Wightman function.  It is worth to note   that  a relevant quantity is not the response function (\ref{Res}) but the rate of  response function given by
\begin{equation} \label{}
\dot{F}(\omega) = \int^{\infty}_{-\infty}d\Delta \tau~ e^{-i\Delta\tau} G^{+}\Big(x(\tau)-x(\tau') \Big)\,.
\end{equation}

Using the Unruh-DeWitt detector method, it was shown  that the Unruh effect is significantly changed by the non-locality~\cite{Nicolini:2009dr}.  This change is interpreted as a kind of the  UV-IR  mixing phenomena  and the entangling aspect is destroyed by the non-locality.
Here, we point out that  this result is inconsistent with the section two in~\cite{Kajuri:2017jmy}. The authors in~\cite{Kajuri:2017jmy}  argued that the point-interacting Unruh-DeWitt detector is not adequate for describing the non-local theory and thus, the results obtained from the detector are not physically meaningful in non-local theories.
However, it is worth to remember  that the Unruh-DeWitt detector method is taken in the inertial frame only, without resorting to the Rindler-Fulling or non-inertial frame quantization scheme. Therefore, the claims for the inadequacy of the Unruh-DeWitt detector seem misleading because it was based on the standard inertial frame quantization.
Moreover, all the information about non-locality or equivalently non point-interacting Unruh-DeWitt detector is already defined in the theory that consists of the Lagrangians $\mathcal{L}$ and $L_{\rm int}$.

In the following, we revisit the Unruh-DeWitt detector method and we prove that the results are completely consistent with those given in section two of~\cite{Kajuri:2017jmy}.
Now, let us first follow the steps to obtain the position-space propagator given in~\cite{Nicolini:2009dr}.
It is straightforward to obtain the Feynman propagator in the momentum space directly from the non-local action
\begin{equation} \label{}
\tilde{\Delta}_{F}(p) = \frac{e^{-\frac{\ell^{2}}{2}p^{2}}}{p^{2}-i\epsilon}\,.
\end{equation}
Since there are some convergence issues in the direct computation in the Lorentzian signature, we take  the Euclidean signature version to compute the propagator.
 We may avoid this signature issue using the form factor $\exp[ - \ell^4\Box^2]$, but we focus on our example for simplicity in the presentation, since we can eventually make an analytic continuation of the amplitudes.
Notice that we rotated $p_0 = i p_E$ in Eq.(11), while below in Eqs. (14) and  (15), we  rotate $x_0 = - i t_E$ only.

Introducing  the Schwinger parametrization of the  propagator
\[
\frac{e^{-\frac{\ell^{2}}{2}p^{2}}}{p^{2}-i\epsilon} = \int^{\infty}_{\frac{\ell^{2}}{2}} ds~ e^{-sp^{2}}\,,
\]
one obtains the non-local propagator in the position-space~\cite{Nicolini:2009dr}
\begin{equation} \label{prop}
\Delta_{F}(x) = \frac{1}{4\pi^{2}x^{2}}\Big(1- e^{-\frac{x^{2}}{2\ell^{2}}}\Big)\,.
\end{equation}

One may be tempting to read off the Wightman function $G^{+}(x)$ from the Feynman propagator $\Delta_{F}(x)$ because these  are related to each other\footnote{See for example~\cite{Papadodimas:2012aq}.
 Our convention for the propagator: $-i \Delta^{(0)}_{F}(x-x') = \langle 0| T\{\phi(x)\phi(x') \} |0 \rangle$.}.
For example, in the local theory (\ref{LST}), the relation is given by
\begin{equation} \label{FWR}
\Delta^{(0)}_{F}(x)= \Theta(t_{E})G^{+}_{(0)}(x) +\Theta(-t_{E})G^{-}_{(0)}(x)\,,
\end{equation}
where $\Delta^{(0)}_{F}(x)$ and $G^{\pm}_{(0)}(x)$ denote the Feynman propagator and the positive (negative)-frequency Wightman function in the local case, respectively.
 Explicitly,  the (Euclidean) Wightman functions can be found through the Wick rotation to be
\begin{align}   \label{Wightman1}
G^{+}_{(0)} (x) &= \int \frac{d^{3}\pmb{p}}{(2\pi)^{3}} \frac{1}{2\omega_{\pmb{p}}} e^{-\omega_{\pmb{p}}t_{E} + i \pmb{p}\cdot \pmb{x}}\,,  \\
\label{Wightman2}G^{-}_{(0)} (x)& = \int \frac{d^{3}\pmb{p}}{(2\pi)^{3}} \frac{1}{2\omega_{\pmb{p}}} e^{\omega_{\pmb{p}}t_{E} - i \pmb{p}\cdot \pmb{x}}
\end{align}
with  $\omega_{\pmb{p}} \equiv \sqrt{\pmb{p}^{2}}$.
  Adopting the same derivation of the Wightman function from the propagator in the non-local case, one arrives at the misleading conclusion that there are huge changes in the Unruh effect~\cite{Nicolini:2009dr}.
   Technically, this huge difference came from the disappearance of the pole in the position-space Feynman propagator~(\ref{prop}) by sending $x\rightarrow 0$.
  It is meaningful to note  that the disappearance of the pole in the position space is different from the one in the  local case ($\ell=0$), but 
it induces a wrong derivation of the Wightman function.

Now, we are in a position to  present  the robustness of the Unruh-DeWitt detector method.
Contrary to the local theory, the non-local  Feynman propagator could not be related to the time ordering of two fields. Indeed, the direct computation leads to
\begin{align}   \label{Fprop}
\Delta_{F}(x)  &= \int \frac{d^{4}p}{(2\pi)^{4}}\tilde{\Delta}_{F}(p) e^{ip\cdot x} =  \int \frac{d^{4}p}{(2\pi)^{4}} \frac{e^{-\frac{\ell^{2}}{2}p^{2}}}{p^{2}}e^{ip\cdot x} \nn \\
& =  e^{\frac{\ell^{2}}{2}\Box_{x}}\bigg[\int \frac{d^{4}p}{(2\pi)^{4}} \frac{1}{p^{2}}e^{ip\cdot x}\bigg] = e^{\frac{\ell^{2}}{2}\Box_{x}}\Delta^{(0)}_{F}(x) \,,
\end{align}
where we  drop the $\epsilon$-prescription because we work in the Euclidean space temporary. The non-local operator  $e^{\frac{\ell^{2}}{2}\Box_{x}}$ could be represented in terms of the kernel $K$ as
\begin{equation} \label{}
e^{\frac{\ell^{2}}{2}\Box_{x}}\Delta^{(0)}_{F}(x) = \int d^{4}y K(x-y) \Delta^{(0)}_{F}(y)\,, \qquad K(x-y) \equiv \frac{1}{(2\pi \ell^{2})^{2}}~ e^{-\frac{(x-y)^{2}}{2\ell^{2}}}\,.
\end{equation}
Using  the (Euclidean) D'Alambertian $\Box = \p^{2}_{t_{E}} + \nabla^{2}$ together with (\ref{FWR})-(\ref{Wightman2}), one finds  that (\ref{Fprop}) takes the form
\begin{equation} \label{propF}
\Delta_{F}(x) =\int \frac{d^{3}\pmb{p}}{(2\pi)^{3}}\bigg[ \frac{1}{2}\text{Erfc}\Big(\frac{\ell\omega_{\pmb{p}}}{\sqrt{2}}-\frac{t_{E}}{\sqrt{2}\ell}\Big)~ G^{+}_{\pmb{p}}(x) + \frac{1}{2}\text{Erfc}\Big(\frac{\ell\omega_{\pmb{p}}}{\sqrt{2}}+\frac{t_{E}}{\sqrt{2}\ell}\Big)~ G^{-}_{\pmb{p}}(x)\bigg]\,.
\end{equation}
Here the complementary error function $\text{Erfc}(z)$ and $G^{\pm}_{\pmb{p}}(x)$'s are  defined by
\begin{equation} \label{}
\text{Erfc}(z) \equiv 1- \text{Erf}(z) =1- \frac{2}{\sqrt{\pi}}\int^{z}_{0}dt~  e^{-t^{2}}\,,  \qquad
G^{\pm}_{\pmb{p}}(x) \equiv \frac{1}{2\omega_{\pmb{p}}}~ e^{\mp\omega_{\pmb{p}}t_{E} +i\pmb{p}\cdot \pmb{x}}\,.
\end{equation}
We would like to mention  that the final expression (\ref{propF}) was derived in Euclidean signature and there are other contributions in the first and forth quarter of the complex energy plane when rotating  back to Minkowski space. However,  the naive Wick rotation may be incorrect in this case.
In deriving the propagator  $\Delta_{F}(x)$~(\ref{propF}), we have used the two relations:
\begin{align}   \label{}
e^{\frac{\ell^{2}}{2}\p^{2}_{t_{E}}}\Big(\theta(t_{E}) e^{\mp\omega_{\pmb{p}}t_{E}} \Big) &= \int^{\infty}_{-\infty} dt'_{E} \frac{1}{\sqrt{2\pi\ell^{2}}} e^{-\frac{(t_{E}-t'_{E})^{2}}{2\ell^{2}} \mp \omega_{\pmb{p}}t'_{E}}\theta(t'_{E})  \nn \\
&= e^{\frac{\ell^{2}}{2}\omega_{\pmb{p}}^2} e^{\mp \omega_{\pmb{p}}t_{E}}~  \frac{1}{2}\text{Erfc}\Big(\frac{\ell\omega_{\pmb{p}}}{\sqrt{2}}\mp \frac{t_{E}}{\sqrt{2}\ell}\Big)\,, \nn \\
e^{\frac{\ell^{2}}{2}\nabla^{2}}\Big( e^{i\pmb{p}\cdot\pmb{x}} \Big) &= e^{-\frac{\ell^{2}}{2}\omega_{\pmb{p}}^2}  e^{i\pmb{p}\cdot\pmb{x}}\,.
\end{align}
It is important to note  that the propagator (\ref{propF})  could also be derived  by adopting the Schwinger parametrization.

 As a check, we remind that  $\text{Erfc}(z)$ reduces to the $\theta$-function in the limit $\ell\rightarrow 0$ as
\[
\lim_{\ell\rightarrow 0}\frac{1}{2} \text{Erfc}\Big(\frac{\ell\omega_{\pmb{p}}}{\sqrt{2}} \mp \frac{t_E}{\sqrt{2}\ell}\Big) = \theta(\pm t_{E} )\, ,
\]
and the positive-frequency Wightman function takes a simple form
\begin{equation} \label{wfun}
G^{+}(x) = \int \frac{d^{3}\pmb{p}}{(2\pi)^{3}}G^{+}_{\pmb{p}}(x)\,,
\end{equation}
which is  the same form of the standard Wightman function (\ref{Wightman1}) for the local theory.
Note that our  Wightman function (\ref{wfun}) satisfies the required properties for the Wightman function given by~(\ref{requirement1}) and~(\ref{requirement2}). Moreover, it  is  consistent with an  expression  obtained through the canonical quantization in the  section two.  Therefore, we insist  that the Wightman function of the non-local theory (\ref{Our}) should  be read as 
\begin{equation} \label{wln}
G^{\pm}(x)= G^{\pm}_{(0)}(x).
\end{equation}
Furthermore, we argue that  the Unruh effect remains unchanged because the non-local Wightman function takes the same form  as in the $\ell=0$  local case.

Finally, we stress that the consistency of the Feynman propagator ~(\ref{propF}) can be checked again  from the micro-causality violation of  the propagator~\cite{Tomboulis:2015gfa}.  The micro-causality violation states  that the relation between the Feynman propagator and the Wightman function for the non-local case ($\ell\neq0$)  cannot be the same as the one for the local case ($\ell=0$).
Although a relation~(\ref{FWR}) for the local theory is required by the micro-causality, it  does not constrain the case of  the non-local theory.
For this purpose, employing the $\theta$-function representation
\[
\theta(\pm t_{E}) = i\int^{\infty}_{-\infty}\frac{d\omega}{2\pi} \frac{e^{\pm i\omega t_{E}}}{\omega + i\epsilon}\,,
\]
one finds
\[
e^{\frac{\ell^{2}}{2}\p^{2}_{t_{E}}}\Big(\theta(t_{E}) e^{\mp\omega_{\pmb{p}}t_{E}} \Big)  = e^{\frac{\ell^{2}}{2}\omega^{2}_{\pmb{p}}}   e^{\mp\omega_{\pmb{p}}t_{E}} \bigg[i\int^{\infty}_{-\infty}\frac{d\omega}{2\pi} \frac{1}{\omega+i\epsilon} e^{\mp i\omega t_{E}} e^{-\frac{\ell^{2}}{2}\omega^{2}+i\ell^{2}\omega_{\pmb{p}}\omega} \bigg].
\]
This  gives us another representation of the complementary error function as
\begin{equation} \label{}
\text{Erfc}\Big(\frac{\ell\omega_{\pmb{p}}}{\sqrt{2}}\mp \frac{t_{E}}{\sqrt{2}\ell}\Big)= 2i\int^{\infty}_{-\infty}\frac{d\omega}{2\pi} \frac{1}{\omega+i\epsilon} e^{\mp i\omega t_{E}} e^{-\frac{\ell^{2}}{2}\omega^{2}+i\ell^{2}\omega_{\pmb{p}}\omega} \,.
\end{equation}
Making use of  this representation together with the Taylor expansion
\[
F(\omega,\omega_{\pmb{p}}) \equiv  e^{-\frac{\ell^{2}}{2}\omega^{2}+i\ell^{2}\omega_{\pmb{p}}\omega}  = 1 -\frac{\ell^{2}}{2}\omega^{2}+i\ell^{2}\omega_{\pmb{p}}\omega + \cdots,
\]
one can rewrite the  non-local Feynman propagator (\ref{propF}) as
\[
\Delta_{F}(x) = \theta(t_{E})G^{+}_{(0)}(x) + \theta(-t_{E})G^{-}_{(0)}(x) + \Delta_{\rm nc}(x)\,.
\]
Here  $G^{\pm}_{(0)}(x)$'s are the Wightman functions (\ref{Wightman1}) and (\ref{Wightman2}) and  $\Delta_{\rm nc}(x)$ denotes  a micro-causality violating term.
Actually,  $\Delta_{\rm nc}(x)$ is composed of an infinite number of contact terms derived in~\cite{Tomboulis:2015gfa}
\begin{equation} \label{}
\Delta_{\rm nc}(x) = -\sum_{m\ge 1}\frac{i^{m-1}}{m!}\frac{\partial^{m-1} \delta(t_{E})}{\partial t^{m-1}_{E}}\Big[ D^{+}(x) -D^{-}(x)\Big]\,,
\end{equation}
where $D^{\pm}(x)$ are defined by
\begin{equation} \label{}
D^{\pm}(x) = \int \frac{d^{3}\pmb{p}}{(2\pi)^{3}} F^{(m)}(\omega_{\pmb{p}},\omega_{\pmb{p}}) G^{\pm}_{\pmb{p}}(x)\,, \qquad F^{(m)}(\omega_{\pmb{p}},\omega_{\pmb{p}}) \equiv \frac{\partial^{m}}{\partial  \omega^{m}}F(\omega,\omega_{\pmb{p}})\Big|_{\omega=\omega_{\pmb{p}}}\,.
\end{equation}
The presence of $\Delta_{\rm nc}(x)$ implies the  micro-causality violation in the non-local theory. 
\section{Field redefinition method and Universality}
In this section, we would like to confirm  our claims that the Unruh effect remains unchanged  in the model~(\ref{Our}) by using field redefinition method.
For this purpose,  we   consider a massive scalar field coupled to a detector
\begin{equation} \label{massive}
S[\phi] = \int d^{4}x \Big[ -\frac{1}{2}e^{\frac{\ell^{2}}{2}(-\Box +m^{2})}\phi (-\Box +m^{2}) \phi\Big]  + g \int d\tau M(x(\tau))\phi(x(\tau))\,.
\end{equation}
Considering  weak non-locality  only, we can  move the form factor in the interaction term by introducing a field redefinition of
\[
\tilde{\phi}=e^{\frac{\ell^{2}}{4}(-\Box +m^{2})}\phi.
\label{FR}
\]
At quantum level,  the Jacobian of the transformation is just a constant.
The  action takes the form
\begin{equation} \label{}
S[\tilde{\phi}] =\int d^{4}x \Big[ -\frac{1}{2}\tilde{\phi}(-\Box +m^{2}) \tilde{\phi}\Big]  + g \int d\tau M(x(\tau))e^{-\frac{\ell^{2}}{4}(-\Box +m^{2})} \tilde{\phi}(x(\tau)).
\end{equation}
Noting that the propagator is the conventional one  of a local theory and it  satisfies the usual K\"{a}llen-Lehman representation, its form is  given by
\begin{equation} \label{}
G^{+}(x-x') = \int \frac{d^{4}p}{(2\pi)^{3}}e^{ip\cdot(x-x')}\theta(p_{0})\delta(p^{2}+m^{2})\,,
\end{equation}
or making  the dependence of  $\tau$ explicitly as 
\begin{equation} \label{}
G^{+}(x(\tau_{1})-x(\tau_{2})) = \int \frac{d^{4}p}{(2\pi)^{3}}e^{ip\cdot(x(\tau_{1})-x(\tau_{2}))}\theta(p_{0})\delta^{(4)}(p^{2}+m^{2}).
\end{equation}
The detector response function reads as
\begin{equation} \label{}
\tilde{F}(\omega) = \int^{\infty}_{-\infty}d\tau  \int^{\infty}_{-\infty}d\tau'~ e^{-i\omega(\tau-\tau')}e^{-\frac{\ell^{2}}{4}(-\Box_{x(\tau)}+m^{2})} e^{-\frac{\ell^{2}}{4}(-\Box_{x(\tau')}+m^{2})}G^{+}(x(\tau)-x(\tau'))\,.
\end{equation}
The response rate function is also given by the general expression
\begin{eqnarray}   \label{trrf}
&& \hspace{-0.8cm} \dot{\tilde{F}} (\omega) =   \int^{\infty}_{-\infty}d\Delta \tau~ e^{-i\omega\Delta\tau}e^{-\frac{\ell^{2}}{4}(-\Box_{x(\tau)}+m^{2})} e^{-\frac{\ell^{2}}{4}(-\Box_{x(\tau')}+m^{2})}G^{+}(x(\tau)-x(\tau'))\,
\nn \\
&& \hspace{-0.8cm} =  \int^{\infty}_{-\infty}d\Delta \tau~ e^{-i\omega\Delta\tau}e^{-\frac{\ell^{2}}{4}(-\Box_{x(\tau)}+m^{2})} e^{-\frac{\ell^{2}}{4}(-\Box_{x(\tau')}+m^{2})}\int \frac{d^{4}p}{(2\pi)^{3}}e^{ip\cdot(x(\tau)-x(\tau'))}\theta(p_{0})\delta^{(4)}(p^{2}+m^{2})~~~  \nn \\
&& \hspace{-0.8cm} = \int^{\infty}_{-\infty}d\Delta \tau~ e^{-i\omega\Delta\tau}\int \frac{d^{4}p}{(2\pi)^{3}}~  e^{-\frac{\ell^{2}}{4}(p^{2}+m^{2})} e^{-\frac{\ell^{2}}{4}(p^{2}+m^{2})} e^{ip\cdot (x(\tau)-x(\tau'))} \theta(p_{0})\delta^{(4)}(p^{2}+m^{2}) \nn \\
&& \hspace{-0.8cm}=  \int^{\infty}_{-\infty}d\Delta \tau~ e^{-i\omega\Delta\tau} \int \frac{d^{4}p}{(2\pi)^{3}}~e^{ip\cdot (x(\tau)-x(\tau'))} \theta(p_{0})\delta^{(4)}(p^{2}+m^{2})\,.
\end{eqnarray}
Importantly,  we observe that $ \dot{\tilde{F}} (\omega)$ is independent of  the form factor [$e^{\frac{\ell^{2}}{4}(-\Box +m^{2})}$] and gives us the same result of the local theory.
 Therefore, there are no
modifications due to the non-locality. It indicates that the Unruh effect is not modified when implementing  a field redefinition in (\ref{massive}).

We note that the analysis in this section is independent of the specific form factor as long as it is a weakly non-local entire function. 
Indeed, in all the formulas in this section we can replace the Gaussian form factor with the exponential of a general entire function, namely
\[
e^{- \frac{\ell^2}{2} \Box} \quad \rightarrow \quad e^{- \frac{1}{2} H(\Box \ell^2)} \, ,
\]
and nothing change if $H(0) = 0$. $H(\Box)$ could be a polynomial of any entire function, while  we fixed $m^2=0$ for the sake of simplicity.
 Therefore, in this section we proved the universality of the Unruh effect in weakly non-local field theories.

It deserve to be notice that the Jacobian of the field redefinition (\ref{FR}) is just a constant and, therefore, does not affect the scattering amplitudes at any order in the loop perturbative expansion. Moreover, (\ref{FR}) does not change the spectrum of the theory introducing or including extra poles in the propagator.

The non-local field theory is well defined in the path integral formalism and all the interactions are analytically  well-defined operators obtained  when expanding the action in perturbations around the selected vacuum.
Weakly non-local or quasi-polynomial theories share most the properties of local theories with two or higher derivative. It seems that everything is very standard.  
Eventually, it is the canonical formulation that should be further investigated, but actually we do not need it.

\section{Conclusion}
In this work, we have revisited the Unruh effect in the simplest non-local theory (\ref{Our}) that captures all the features of a large class of weakly non-local field theories.
By  considering the non-local model specified by the exponential of the d'Alembertian operator,
we have confirmed that various methods adopted in exploring  the Unruh effect give us the same results of the local scalar theory, unlike to the drastic change found in~\cite{Nicolini:2009dr}. It turns out that the Unruh effect is not modified in the  non-local model~(\ref{Our}) respect to the local field theory (\ref{LST}). Consequently, the Davies-Unruh temperature is unchanged. 

 More concretely, we have revisited the Unruh-DeWitt detector method and we have found  that it leads to the same results as those from the canonical approach. Contrary to the previous claims~\cite{Kajuri:2017jmy}, we have shown that the detector model is robust and can be used to describe  the Unruh effect correctly.
Even though we have focused on a non-local scalar theory of Gaussian exponential type, our main conclusion would hold for any weakly non-local or quasi polynomial theory \cite{Modesto:2011kw}.

Furthermore, we have checked our claim 
by using the field redefinition approach. Indeed, all the scattering amplitudes are invariant under weakly non-local field redefinition that turns the propagator into the local one, but leads to changing the interaction vertexes. We mention that the theories before and after field redefinition are identical at perturbative level and at any order in the loop expansion.

Finally, we would like to point out the reason why there is a significant modification in the  Newtonian potential $V(r)=-GM{\rm Erf}(r/2l)/r$~\cite{Biswas:2011ar, Modesto:2014eta}, whereas  there is no modifications in the Unruh effect.
This can be simply understood by looking at the Feynman propagator and the Wightman function that involve the potential and the Unruh effect respectively.
Indeed, the Feynman propagator $\Delta_{F}(x,x')$ determining the potential  satisfies a Poisson-like equation with a Delta source, which becomes a Gaussian-like source in the non-local case.
On the other hand, the Wightman function determining the Unruh effect  satisfies the Laplace equation without source.
The last statement supports our claim strongly that  the Unruh effect is not modified in the non-local model.

\vskip 1cm
\centerline{\large \bf Acknowledgments}
\vskip0.5cm
{YM was supported by the National Research Foundation of Korea
(NRF) grant funded by the Korea government (MOE) (No. NRF-2017R1A2B4002057). SY was supported by the National Research Foundation of Korea (NRF) funded by the Korea government (MOE) (Nos. NRF-2015R1D1A1A09057057 and  NRF-2017R1A2B4002057).

\newpage

%


\begin{thebibliography}{99}



\bibitem{Woodard:2006nt}
  R.~P.~Woodard,
  ``Avoiding dark energy with $1/r$ modifications of gravity,''
  Lect.\ Notes Phys.\  {\bf 720}, 403 (2007)
  [astro-ph/0601672].

\bibitem{Bombelli:1987aa}
  L.~Bombelli, J.~Lee, D.~Meyer and R.~Sorkin,
  ``Space-Time as a Causal Set,''
  Phys.\ Rev.\ Lett.\  {\bf 59}, 521 (1987).
  doi:10.1103/PhysRevLett.59.521

\bibitem{CaMo2} 
  G.~Calcagni and L.~Modesto,
  ``Nonlocal quantum gravity and M-theory,''
  Phys.\ Rev.\ D {\bf 91}, no. 12, 124059 (2015)
  [arXiv:1404.2137 [hep-th]].

  \bibitem{krasnikov}
  N.~V.~Krasnikov,
  ``Nonlocal Gauge Theories,''
  Theor.\ Math.\ Phys.\  {\bf 73}, 1184 (1987)
  [Teor.\ Mat.\ Fiz.\  {\bf 73}, 235 (1987)].

\bibitem{kuzmin}
  Y.~V.~Kuz'min,
  ``The Convergent Nonlocal Gravitation (in Russian),''
  Sov.\ J.\ Nucl.\ Phys.\  {\bf 50}, 1011 (1989)
  [Yad.\ Fiz.\  {\bf 50}, 1630 (1989)].


\bibitem{tom}
  E.~T.~Tomboulis,
  hep-th/9702146;
%
  E.~T.~Tomboulis,
  ``Renormalization and unitarity in higher derivative and nonlocal gravity theories,''
  Mod.\ Phys.\ Lett.\ A {\bf 30}, 1540005 (2015).

\bibitem{Modesto:2011kw}
  L.~Modesto,
  ``Super-renormalizable Quantum Gravity,''
  Phys.\ Rev.\ D {\bf 86}, 044005 (2012)
  [arXiv:1107.2403 [hep-th]].


\bibitem{Douglas:2001ba}
  M.~R.~Douglas and N.~A.~Nekrasov,
  ``Noncommutative field theory,''
  Rev.\ Mod.\ Phys.\  {\bf 73}, 977 (2001)
  [hep-th/0106048].

\bibitem{Modesto:2014lga}
  L.~Modesto and L.~Rachwal,
  ``Super-renormalizable and finite gravitational theories,''
  Nucl.\ Phys.\ B {\bf 889}, 228 (2014)
  [arXiv:1407.8036 [hep-th]].



\bibitem{Unruh:1976db}
  W.~G.~Unruh,
  ``Notes on black hole evaporation,''
  Phys.\ Rev.\ D {\bf 14}, 870 (1976).


\bibitem{Nicolini:2009dr}
  P.~Nicolini and M.~Rinaldi,
  ``A Minimal length versus the Unruh effect,''
  Phys.\ Lett.\ B {\bf 695}, 303 (2011)
  doi:10.1016/j.physletb.2010.10.051
  [arXiv:0910.2860 [hep-th]].


\bibitem{Kajuri:2017jmy}
  N.~Kajuri,
  ``Unruh Effect in nonlocal field theories,''
  Phys.\ Rev.\ D {\bf 95}, no. 10, 101701 (2017)
  [arXiv:1704.03793 [gr-qc]].


\bibitem{Fulling:1972md}
  S.~A.~Fulling,
  ``Nonuniqueness of canonical field quantization in Riemannian space-time,''
  Phys.\ Rev.\ D {\bf 7}, 2850 (1973).


\bibitem{Barnaby:2007ve}
  N.~Barnaby and N.~Kamran,
  ``Dynamics with infinitely many derivatives: The Initial value problem,''
  JHEP {\bf 0802}, 008 (2008)
  [arXiv:0709.3968 [hep-th]].

\bibitem{Calcagni} 
  G.~Calcagni and G.~Nardelli,
  ``Non-local gravity and the diffusion equation,''
  Phys.\ Rev.\ D {\bf 82}, 123518 (2010)
  [arXiv:1004.5144 [hep-th]].
  G.~Calcagni and G.~Nardelli,
  ``String theory as a diffusing system,''
  JHEP {\bf 1002}, 093 (2010)
  [arXiv:0910.2160 [hep-th]];
  G.~Calcagni, M.~Montobbio and G.~Nardelli,
  ``Localization of nonlocal theories,''
  Phys.\ Lett.\ B {\bf 662}, 285 (2008)
  [arXiv:0712.2237 [hep-th]].



\bibitem{Barci:1995ad}
  D.~G.~Barci, L.~E.~Oxman and M.~Rocca,
  ``Canonical quantization of nonlocal field equations,''
  Int.\ J.\ Mod.\ Phys.\ A {\bf 11}, 2111 (1996)
  [hep-th/9503101].


\bibitem{Aslanbeigi:2014zva}
  S.~Aslanbeigi, M.~Saravani and R.~D.~Sorkin,
  ``Generalized causal set d`Alembertians,''
  JHEP {\bf 1406}, 024 (2014)
  doi:10.1007/JHEP06(2014)024
  [arXiv:1403.1622 [hep-th]].


\bibitem{Weinberg:1995mt}
  S.~Weinberg,
  ``The Quantum theory of fields. Vol. 1: Foundations''.


\bibitem{Wald:1995yp}
  R.~M.~Wald,
  ``Quantum Field Theory in Curved Space-Time and Black Hole Thermodynamics''.


\bibitem{Takagi:1986kn}
  S.~Takagi,
  ``Vacuum noise and stress induced by uniform accelerator: Hawking-Unruh effect in Rindler manifold of arbitrary dimensions,''
  Prog.\ Theor.\ Phys.\ Suppl.\  {\bf 88}, 1 (1986).


\bibitem{Crispino:2007eb}
  L.~C.~B.~Crispino, A.~Higuchi and G.~E.~A.~Matsas,
  ``The Unruh effect and its applications,''
  Rev.\ Mod.\ Phys.\  {\bf 80}, 787 (2008)
  [arXiv:0710.5373 [gr-qc]].


\bibitem{Unruh:1983ms}
  W.~G.~Unruh and R.~M.~Wald,
  ``What happens when an accelerating observer detects a Rindler particle,''
  Phys.\ Rev.\ D {\bf 29}, 1047 (1984).


\bibitem{Papadodimas:2012aq}
  K.~Papadodimas and S.~Raju,
  ``An Infalling Observer in AdS/CFT,''
  JHEP {\bf 1310}, 212 (2013)
  doi:10.1007/JHEP10(2013)212
  [arXiv:1211.6767 [hep-th]].

\bibitem{Tomboulis:2015gfa}
  E.~T.~Tomboulis,
  ``Nonlocal and quasilocal field theories,''
  Phys.\ Rev.\ D {\bf 92}, no. 12, 125037 (2015)
  [arXiv:1507.00981 [hep-th]].

\bibitem{Biswas:2011ar}
  T.~Biswas, E.~Gerwick, T.~Koivisto and A.~Mazumdar,
  ``Towards singularity and ghost free theories of gravity,''
  Phys.\ Rev.\ Lett.\  {\bf 108}, 031101 (2012)
  [arXiv:1110.5249 [gr-qc]].

\bibitem{Modesto:2014eta}
  L.~Modesto, T.~de Paula Netto and I.~L.~Shapiro,
  JHEP {\bf 1504}, 098 (2015)
  [arXiv:1412.0740 [hep-th]].


\end{thebibliography}
%


\end{document}